\documentclass{jpsj-suppl}
\usepackage{txfonts} %Please comment out this line unless the txfonts package is availabe in your LaTeX system.

\usepackage{hyperref}

\def\ba{\begin{eqnarray}}
\def\ea{\end{eqnarray}}

\title{Empirical parametrizations of the resonance 
amplitudes based on the Siegert's theorem}

\author{G.~Ramalho}

\inst{International Institute of Physics, 
Federal University of Rio Grande do Norte, 
Campus Lagoa Nova - Anel Vi\'ario da UFRN, 
Lagoa Nova, Natal-RN, 59070-405, Brazil}

\email{gilberto.ramalho@iip.ufrn.br}

\recdate{....}

\abst{We present parametrizations 
of the $\gamma^\ast N \to N(1535)1/2^-$, $\gamma^\ast N \to N(1520)3/2^-$
and  $\gamma^\ast N \to \Delta(1232)3/2^+$ transition amplitudes 
that are compatible with the analytic constraints 
at the pseudothreshold (Siegert's theorem).
The presented parametrizations also provide a fair 
description of the experimental data.
For the case of the $\gamma^\ast N \to \Delta(1232)3/2^+$ 
transition, we discuss how the 
pion cloud parametrizations 
of the electric and the Coulomb quadrupole form factors 
can be adjusted according to the Siegert's theorem.}

\kword{nucleon resonances, transition amplitudes, Siegert's theorem} 
\begin{document}
\maketitle

\vspace{-.7cm}

\section{Introduction}

The electroexcitation of the nucleon resonances ($\gamma^\ast N \to N^\ast$)
is experimentally restricted to the spacelike region when 
the photon momentum transfer $q^2$ is negative
($Q^2 = - q^2 > 0$)~\cite{NSTAR}.
The timelike region is delimited by the interval 
$-(M_R-M)^2 < Q^2 < 0$, where $M$ and  $M_R$ are respectively 
the nucleon and the resonance  masses.
Although the pseudothreshold is close to 
the  photon point $Q^2=0$  
for light resonances, 
the timelike region cannot be directly accessed.
The timelike region however imposes 
constraints on the analytic form 
of the transition amplitudes and form factors
in the limit where the photon three-momentum ${\bf q}$
vanishes and $Q^2= -(M_R-M)^2$
(pseudothreshold limit).
When the transition current is defined using  
the general on-mass-shell gauge invariant structure 
one can define {\it elementary} form factors,
such as the case of the Dirac and Pauli form factors 
for transitions to spin 1/2 resonances.
The requirement that those form factors are 
free of kinematic singularities at the pseudothreshold
implies that two or more form factors are correlated 
at the pseudothreshold~\cite{Bjorken66,Devenish76}.
An implication of those constraints 
is the relation between the 
electric ($E$) and scalar ($S$) amplitudes 
expressed by $E \propto S/|{\bf q}|$, 
when $|{\bf q}| \to 0$, which is known as the Siegert's 
theorem~\cite{Drechsel92,Drechsel2007,Tiator15}.
In this work we discuss in particular 
the Siegert's theorem for the 
$\gamma^\ast N \to N(1535)1/2^-$, $\gamma^\ast N \to N(1520)3/2^-$
and  $\gamma^\ast N \to \Delta(1232)3/2^+$ 
transitions~\cite{Siegert,SiegertD,SiegertQuad}.

\section{$\gamma^\ast N \to N(1535)1/2^-$}

In the case of the $\gamma^\ast N \to N(1535)1/2^-$ transition one can 
concludes that the longitudinal amplitude ($A_{1/2}$) 
and the scalar amplitude ($S_{1/2}$) can 
be expressed at pseudothreshold limit as~\cite{Devenish76,Siegert}
\ba
A_{1/2} = 2 b \tilde F_1, \hspace{1.5cm} 
S_{1/2}= \frac{\sqrt{2}}{M_R-M} \tilde F_1 |{\bf q}|,
\label{eqS11}
\ea
where $b= e \sqrt{\frac{Q^2 + (M_R -M)^2}{8 M (M_R^2-M^2)}}$
($e$ is the elementary charge) and $\tilde F_1 = F_1 + \eta F_2$
is a combination of the Dirac and Pauli form factors
with $\eta= \frac{M_R -M}{M_R + M}$~\cite{Siegert}.
In addition one concludes that at the pseudothreshold,
$\tilde F_1$ is finite and the functions  
$F_1, F_2$ are free from kinematic singularities. 
The direct consequence of Eq.~(\ref{eqS11}) 
is $A_{1/2} = \lambda S_{1/2}/|{\bf q}|$ with 
$\lambda = \sqrt{2} (M_R-M)$, which 
is equivalent to the Siegert's theorem, 
since $E \propto A_{1/2}$.
The result $S_{1/2} \propto |{\bf q}|$ 
in particular can be interpreted as a consequence 
of the orthogonality between the nucleon and the 
$N(1535)1/2^-$ states~\cite{Siegert}. 

A parametrization of the amplitude $S_{1/2}$ 
compatible with the Siegert's theorem is 
presented 
%in Fig.~\ref{figN1535} 
in Fig.~1 
(left panel)
with the label MAID-SG~\cite{Siegert}. 
This parametrization differs from the MAID2007 
parametrization~\cite{Drechsel2007} 
which fails to describe the Siegert's theorem.
The constraint from the Siegert's theorem 
is responsible for the  inflection of the 
function $S_{1/2}$ near the pseudothreshold
in the MAID-SG parametrization. 
In the right panel we present the results 
for the kinematic-free form factors $F_1$ and $F_2$.
More details can be found in Ref.~\cite{Siegert}. 

\begin{figure*}[h]%[tbh]
%\vspace{.2cm}
\centerline{
\mbox{
\includegraphics[width=2.3in]{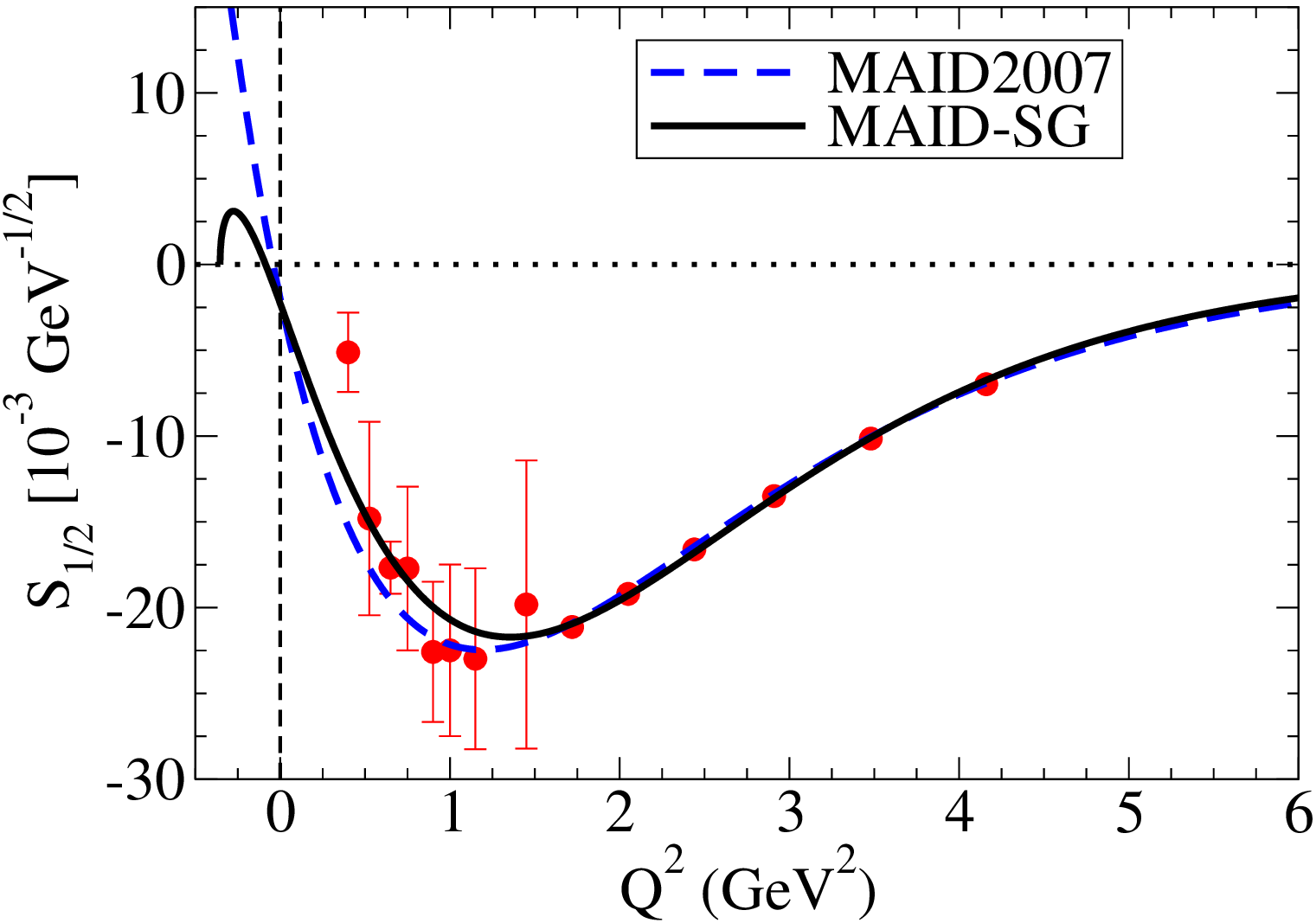} \hspace{1.5cm}
\includegraphics[width=2.1in]{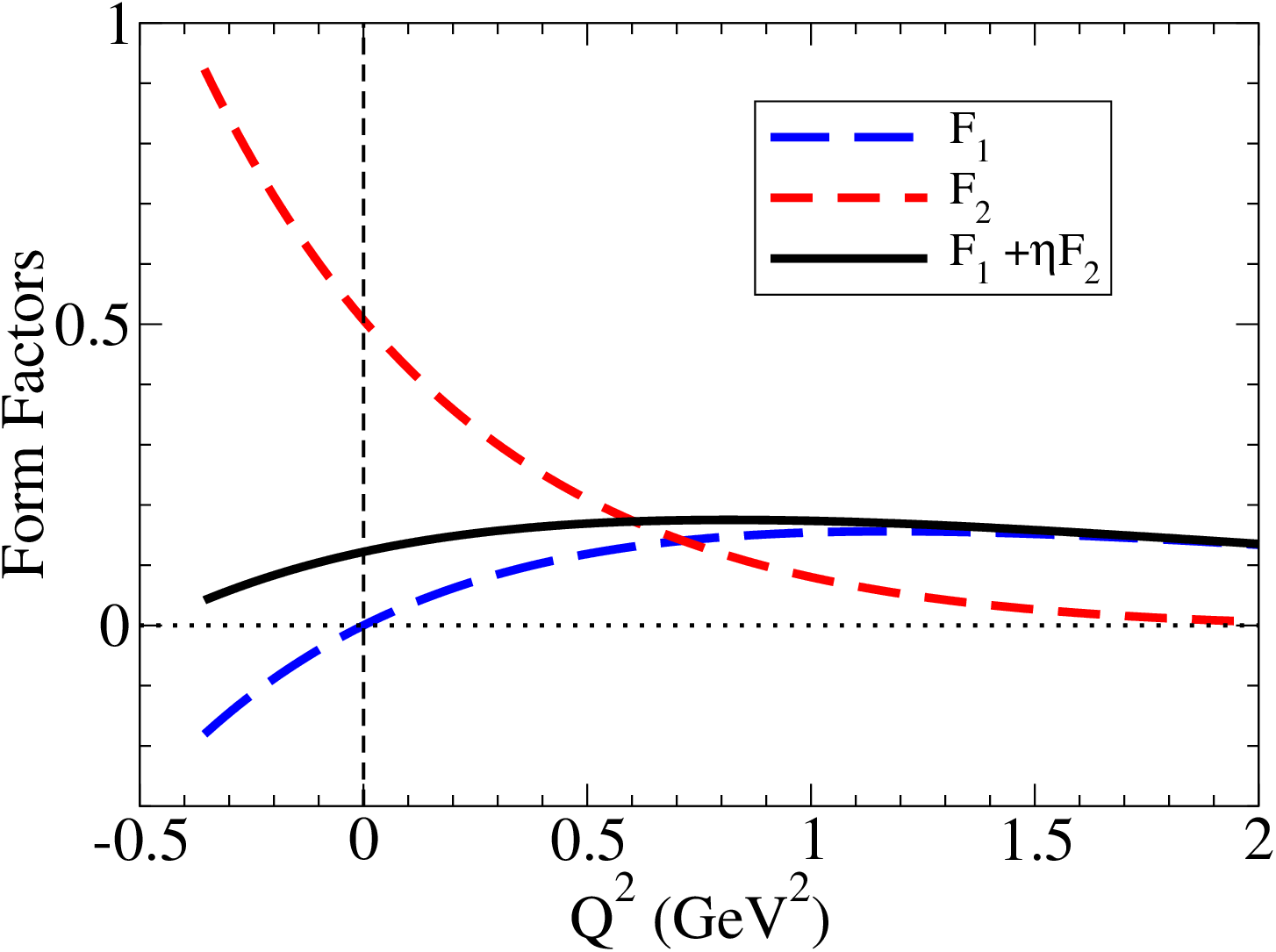} }}
%\caption{$\gamma^\ast N \to N(1535)1/2^-$ transition. At the left: Amplitude $S_{1/2}$, comparison between a parametrization compatible with the Siegert's theorem (MAID-SG) and MAID2007. At the right: Representation of the Dirac ($F_1$)  and Pauli ($F_2$) form factors (MAID-SG). Data from Ref.~\cite{MokeevDatabase}.}
\label{figN1535}
\end{figure*}
\vspace{-.7cm}
\noindent
{\bf Fig.~1.}
{$\gamma^\ast N \to N(1535)1/2^-$ transition.
At the left: Amplitude $S_{1/2}$, comparison between 
a parametrization compatible with the Siegert's theorem (MAID-SG) 
and MAID2007.
At the right: Representation of the Dirac ($F_1$) 
and Pauli ($F_2$) form factors (MAID-SG).
Data from Ref.~\cite{MokeevDatabase}.}

%\vspace{-1.17cm}

\begin{figure*}[h] %[tbh]
\label{figN1520}
%\vspace{-.5cm}
\centerline{
\mbox{
\includegraphics[width=2.4in]{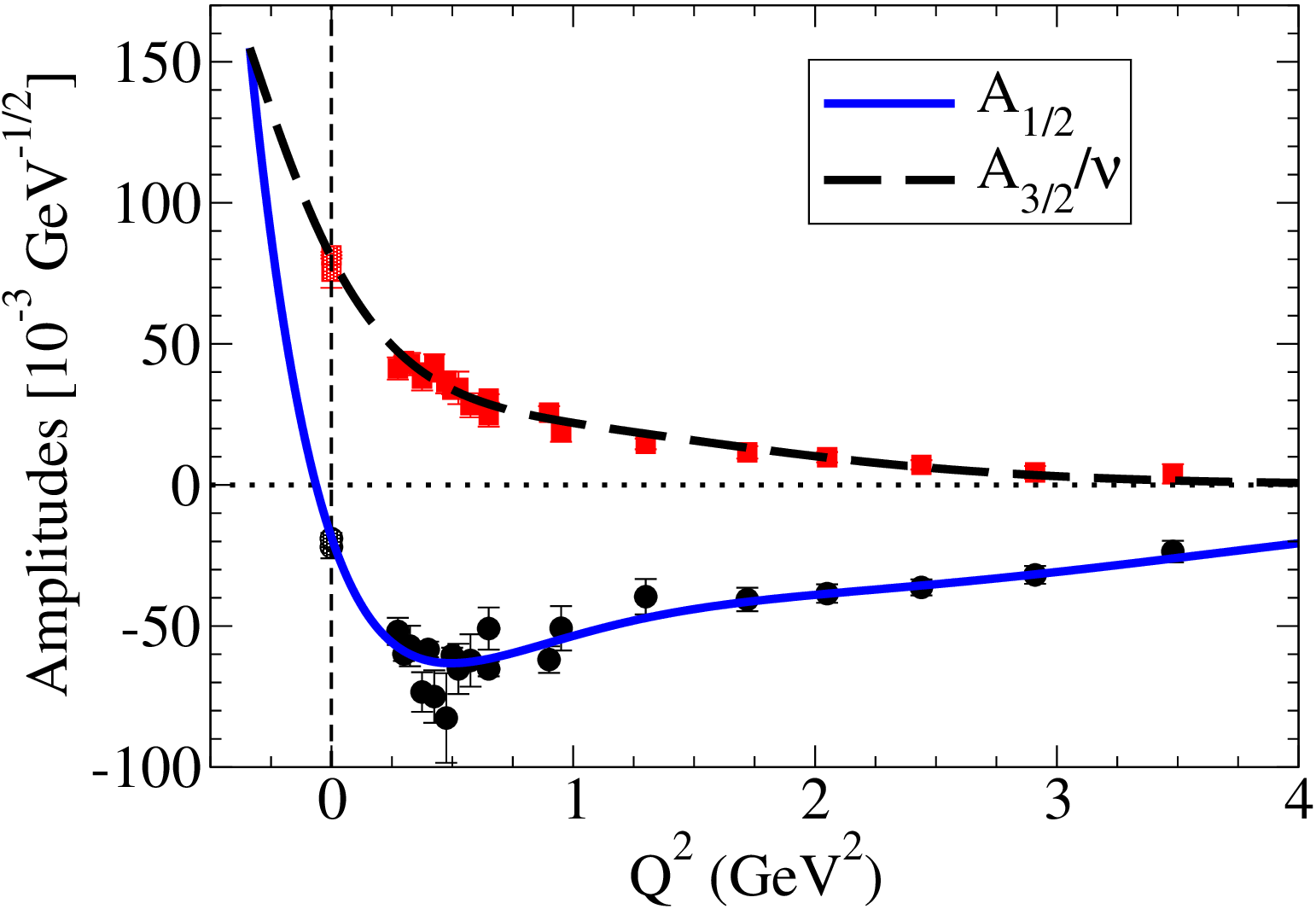}  \hspace{1.5cm}
\includegraphics[width=2.2in]{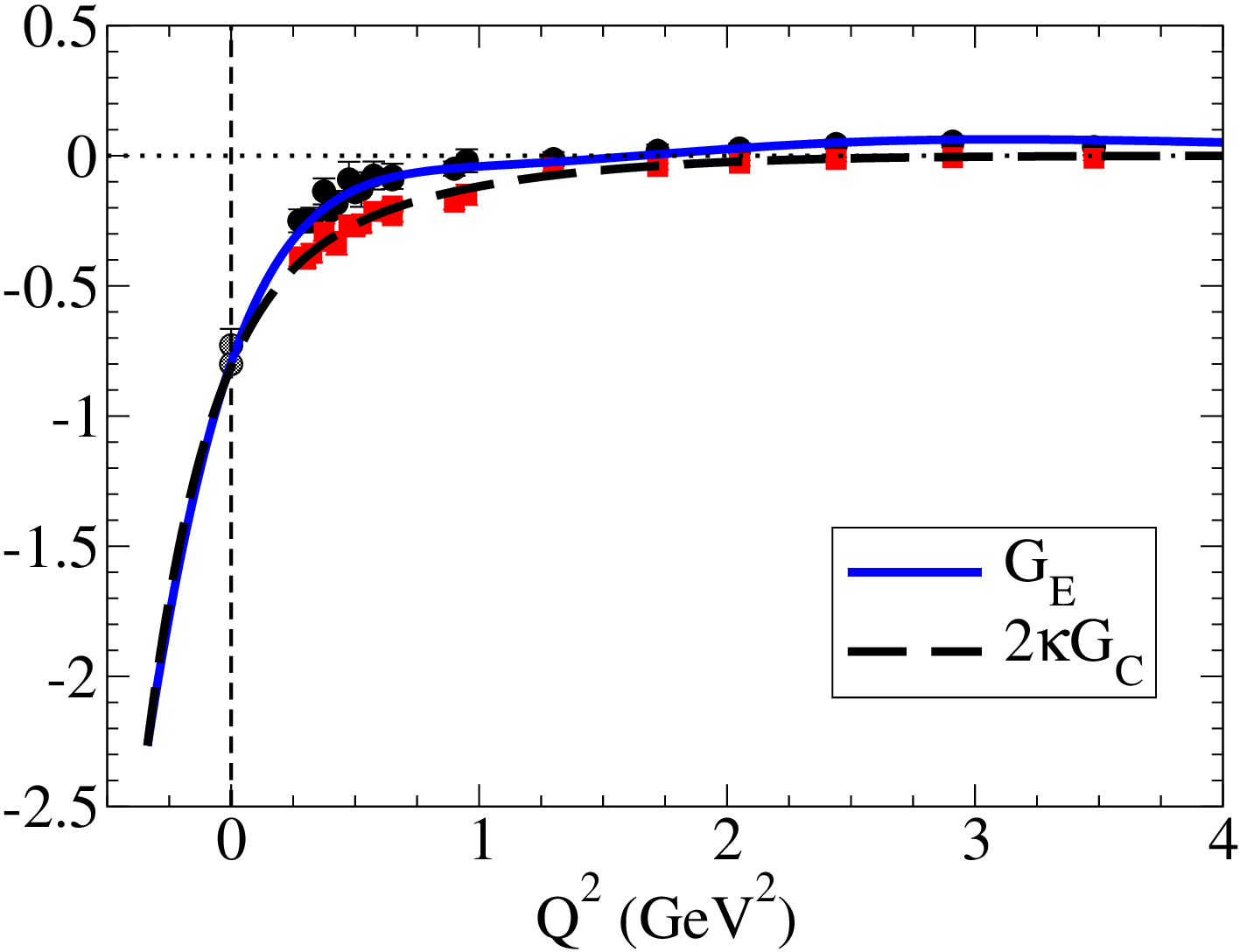}
}}
%\caption{$\gamma^\ast N \to N(1520)3/2^-$ transition. At the left: Comparison between transverse amplitudes $A_{1/2}$  and $A_{3/2}$ ($\nu =\sqrt{3}$).  At the right: Comparison between the form factors $G_E$ and $\kappa G_C$, with $\kappa= \frac{M_R -M}{2 M_R}$. Data from Ref.~\cite{MokeevDatabase}.}
\end{figure*}
\vspace{-.7cm}
\noindent
{\bf Fig.~2.}
{$\gamma^\ast N \to N(1520)3/2^-$ transition.
At the left: Comparison between transverse amplitudes $A_{1/2}$ 
and $A_{3/2}$ ($\nu =\sqrt{3}$). 
At the right: Comparison between the form factors 
$G_E$ and $\kappa G_C$, with $\kappa= \frac{M_R -M}{2 M_R}$.
Data from Ref.~\cite{MokeevDatabase}.}

\section{$\gamma^\ast N \to N(1520)3/2^-$}

%\vspace{-.17cm}

In the case of the $\gamma^\ast N \to N(1520)3/2^-$ transition 
the analytic structure  (singularity-free form factors) 
requires the following dependence of the amplitudes 
near the pseudothreshold 
\ba
M_{2-}= {\cal O}(|{\bf q}|^2), \hspace{.8cm}
E_{2-}= {\cal O}(1), \hspace{1.cm}
S_{1/2} = {\cal O}(|{\bf q}|),
\label{eqN1520}
\ea
where $M_{2-}$, $E_{2-}$ and $S_{1/2}$ are 
the magnetic, electric and scalar amplitudes~\cite{Devenish76,Drechsel92}.
When expressed in terms of the transverse amplitudes,
$A_{1/2}$ and $A_{3/2}$, one can conclude that 
the first condition from Eq.~(\ref{eqN1520}) imply that 
$A_{1/2}= A_{3/2}/\sqrt{3}$.
In addition, at the pseudothreshold
the last two amplitudes 
are related by  
$\frac{1}{2} E_{2-} = \lambda S_{1/2}$~\cite{Devenish76,SiegertD}.
Using the parametrizations for $A_{1/2}, A_{3/2}$ 
and $S_{1/2}$ one can calculate 
the transition form factors $G_M,G_E$ and $G_C$.
At the pseudothreshold $G_E= \frac{M_R-M}{M_R} G_C$.
Parametrizations of the data compatible 
with the previous two conditions are presented
%in Fig.~\ref{figN1520}. 
in Fig.~2.
In the figure
we can see that both conditions are satisfied
in the lower limit of $Q^2$.
More details can be found in Ref.~\cite{SiegertD}. 

%\vspace{-.7cm}

\section{$\gamma^\ast N \to \Delta(1232)3/2^+$}

The $\gamma^\ast N \to \Delta(1232)3/2^+$ transition 
can be characterized by the amplitudes 
$A_{1/2}, A_{3/2}$ and $S_{1/2}$, or alternatively 
by the magnetic ($G_M$) electric ($G_E$) 
and Coulomb ($G_C$) form factors.
The pseudothreshold limit implies that
\ba
\frac{E_{1+}}{|{\bf q}|} = \lambda \frac{S_{1/2}}{|{\bf q}|^2},
\hspace{1cm}
\lambda = \sqrt{2}(M_R -M),
\label{eqDelta}
\ea 
which is equivalent to the relation 
between electric and Coulomb quadrupole 
form factors: $G_E = \frac{M_R -M}{2 M_R} G_C$~\cite{Jones73}.
A parametrization of the data (MAID-SG2)~\cite{SiegertD} 
compatible with the previous condition 
is presented in the left panel of 
%Fig.~\ref{figD1232-1}.
Fig.~3.
The present form of the Siegert's theorem, Eq.~(\ref{eqDelta}) 
differs from the more common form 
$E_{1+} = \lambda S_{1/2}/|{\bf q}|$~\cite{Drechsel2007,Tiator15}
by a factor of $1/|{\bf q}|$.
This factor is however necessary in order 
to obtain the correct relation between form factors at 
the pseudothreshold~\cite{SiegertD}.

In the central and right panel of 
%Fig.~\ref{figD1232-1}
Fig.~3
we present the ratios $R_{EM}, R_{SM}$ associated with 
the form factor data for $G_E$ and $G_C$ 
corresponding to the MAID-SG2 parametrization,
for larger values of $Q^2$.
The MAID-SG2 parametrization is based on rational functions of $Q^2$,
inducing the falloffs $G_E \propto 1/Q^4$, 
$G_C \propto 1/Q^6$ for very large $Q^2$,
as expected from pQCD~\cite{SiegertD}.

\begin{figure*}[tbh]
\centerline{
\mbox{
\includegraphics[width=1.9in]{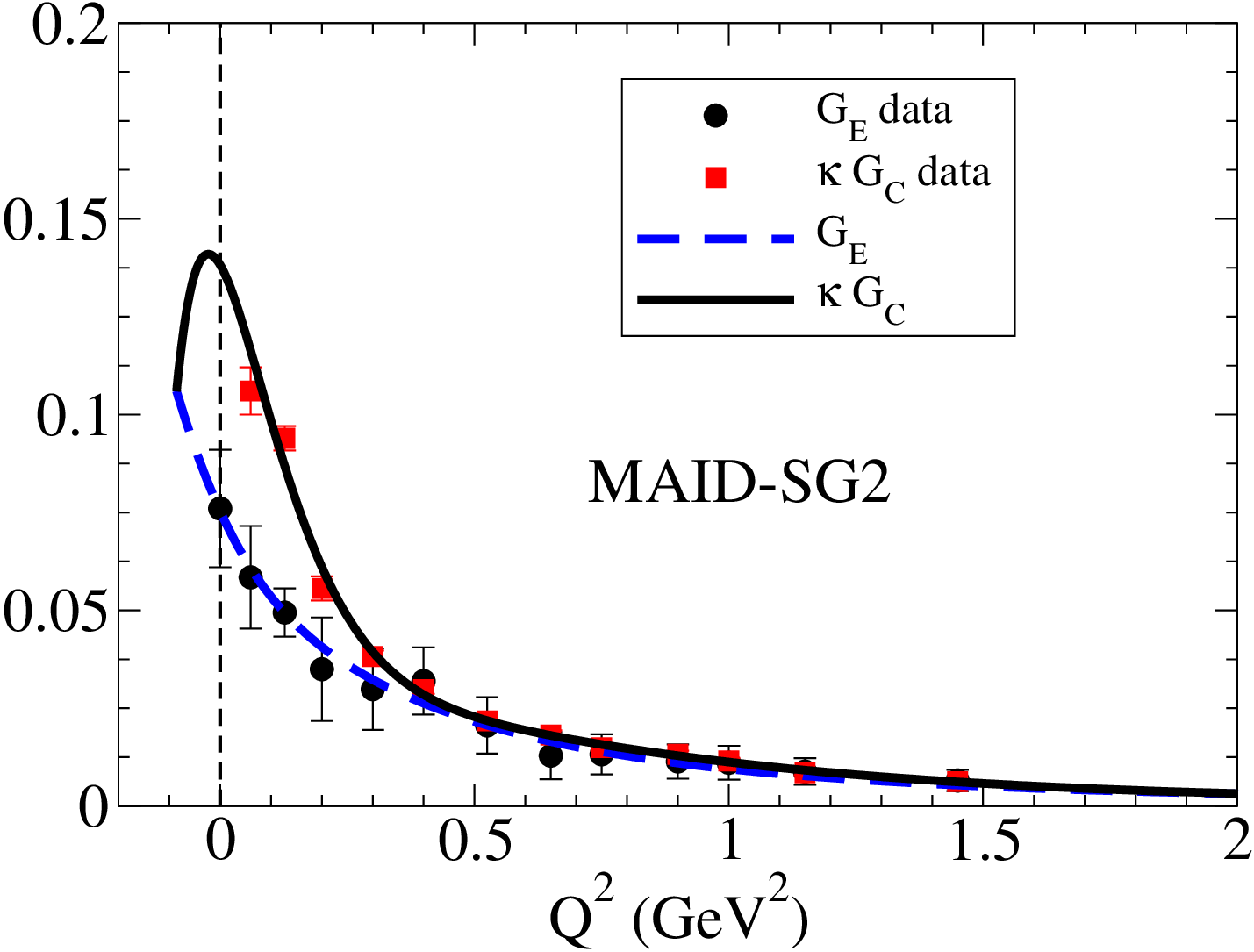} \hspace{.2cm}
\includegraphics[width=1.9in]{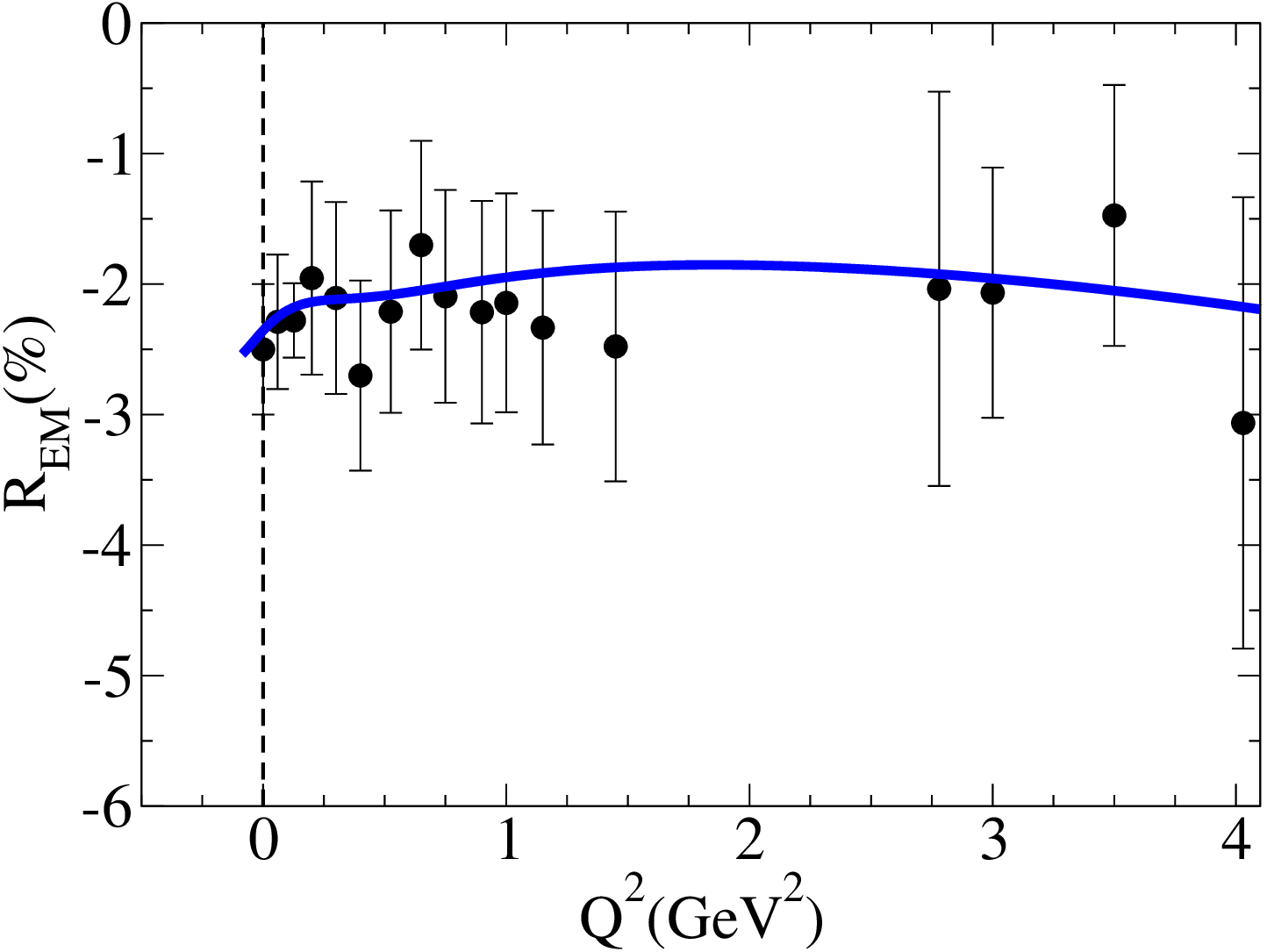} \hspace{.2cm}
\includegraphics[width=1.9in]{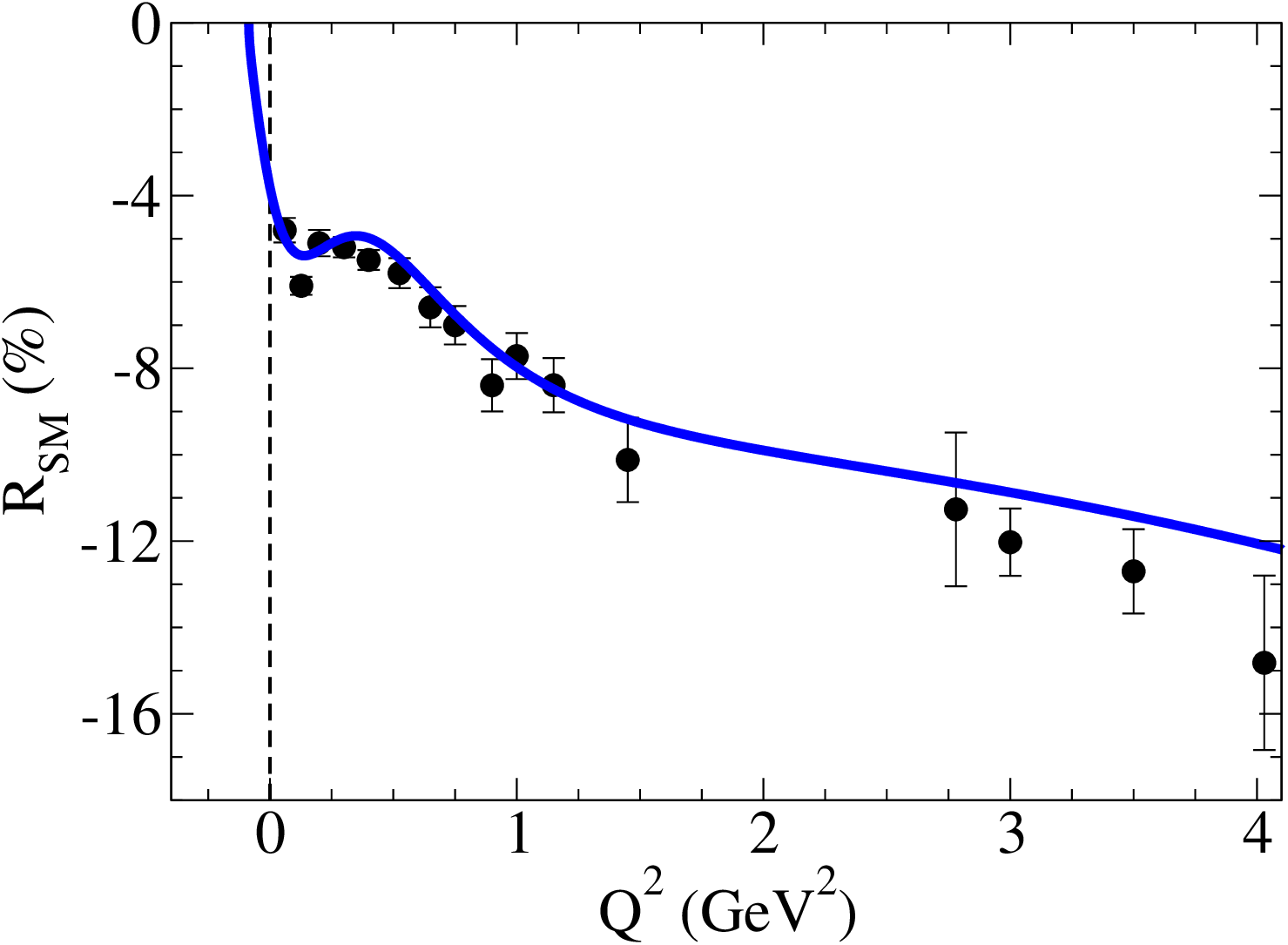} }}
%\caption{$\gamma^\ast N \to \Delta(1232)3/2^+$ transition. At the left: Functions $G_E$ and $\kappa G_C$ with  $\kappa=  \frac{M_R -M}{2 M_R}$. At the center: Results for $R_{EM} \equiv - \frac{G_E}{G_M}$. At the right: Results for $R_{SM} \equiv - \frac{|{\bf q}|}{2 M_R}\frac{G_E}{G_M}$. Data from Ref.~\cite{MokeevDatabase}.}
\label{figD1232-1}
\end{figure*}
\vspace{-.7cm}
\noindent
{\bf Fig.~3.}
{$\gamma^\ast N \to \Delta(1232)3/2^+$ transition.
At the left: Functions $G_E$ and $\kappa G_C$ with  
$\kappa=  \frac{M_R -M}{2 M_R}$.
At the center:
Results for $R_{EM} \equiv - \frac{G_E}{G_M}$.
At the right: 
Results for $R_{SM} \equiv - \frac{|{\bf q}|}{2 M_R}\frac{G_E}{G_M}$.
Data from Ref.~\cite{MokeevDatabase}.}

\section{$\gamma^\ast N \to \Delta(1232)3/2^+$ -- 
$G_E$ and $G_C$ parametrizations}

The Siegert's theorem has been discussed in the literature 
in the context of constituent quark 
models~\cite{SiegertD,Buchmann98}.
The main conclusion is that a consistent description 
of the transition current requires the inclusion 
of processes beyond the impulse approximation at the quark level.
This means that two-body exchange currents
that include quark-antiquark states
are important for the description of the form factors 
near the pseudothreshold and at low $Q^2$~\cite{Buchmann98}.
Those contributions are also refereed to as meson cloud contributions. 

When we consider the one-body level 
(impulse approximation) the 
Siegert's theorem is trivially 
satisfied since the baryon wave functions 
based on valence quark contributions 
lead to vanishing form factors at the 
pseudothreshold as a consequence of 
the orthogonality between states.
For the $\gamma^\ast N \to \Delta(1232)3/2^+$ this 
was shown explicitly in Ref.~\cite{SiegertQuad} 
based on a covariant quark model~\cite{Nucleon,NDeltaD,LatticeD}.

The problem becomes more complex when we consider
contributions associated with the meson cloud 
(beyond impulse approximation).
Considering the $\gamma^\ast N \to \Delta(1232)3/2^+$ 
transition, there are parametrizations 
of the pion cloud contributions 
for the quadrupole form factors 
$G_E$ and $G_C$ derived in the large-$N_c$ limit 
that relate those form factors 
with the electric form factor of 
the neutron ($G_{En}$)~\cite{Buchmann,Pascalutsa07a}
\ba
G_E^\pi (Q^2)= \left( 
\frac{M}{M_\Delta}\right)^{3/2}
\frac{M_\Delta^2- M^2}{2 \sqrt{2}} 
\tilde G_{En}(Q^2),
\hspace{1cm}
G_C^\pi (Q^2)= \left( 
\frac{M}{M_\Delta}\right)^{1/2}
\sqrt{2}M M_\Delta
\tilde G_{En}(Q^2),
\label{eqLargeNc}
\ea
where $M_\Delta$ is the mass of the $\Delta$
and $\tilde G_{En}= G_{En}(Q^2)/Q^2$.
One can however show  that Eqs.~(\ref{eqLargeNc}) 
are not compatible with the Siegert's theorem.
Using $G_{En} \simeq - \frac{1}{6} r_n^2 Q^2$, 
for small values of $Q^2$,
 one concludes that 
$G_E^\pi - \kappa G_C^\pi = {\cal O}(1/N_c^2)$
at the pseudothreshold.
A better approximation can be obtained 
by modifying $G_E^\pi$ with a factor $1/(1 + Q^2/(M_\Delta^2-M^2))$,
which corresponds to a relative correction of $1/N_c^2$ 
at the pseudothreshold.
The value of $G_E^\pi$ at $Q^2=0$ is however unchanged. 
Using the new form for $G_E^\pi$, 
one has $G_E^\pi - \kappa G_C^\pi = {\cal O}(1/N_c^4)$
at the pseudothreshold, which corresponds 
to a significant improvement comparative
to the parametrizations from Eqs.~(\ref{eqLargeNc})~\cite{SiegertQuad}.

Combining the new parametrizations for 
$G_E^\pi$ and $G_C^\pi$ with a contributions 
from the valence quark core estimated 
by the covariant spectator quark model 
extrapolated from lattice QCD data~\cite{LatticeD,Alexandrou08},
one obtains an estimate 
of the form factors 
$G_E$ and $G_C$ compatible with the Siegert's theorem 
up to $1/N_c^4$ corrections.
The results are presented in 
%Fig.~\ref{figD1232-2}.
Fig.~4.
The results are in excellent agreement with the data.

\begin{figure*}[tbh]
\centerline{
\mbox{
\includegraphics[width=2.15in]{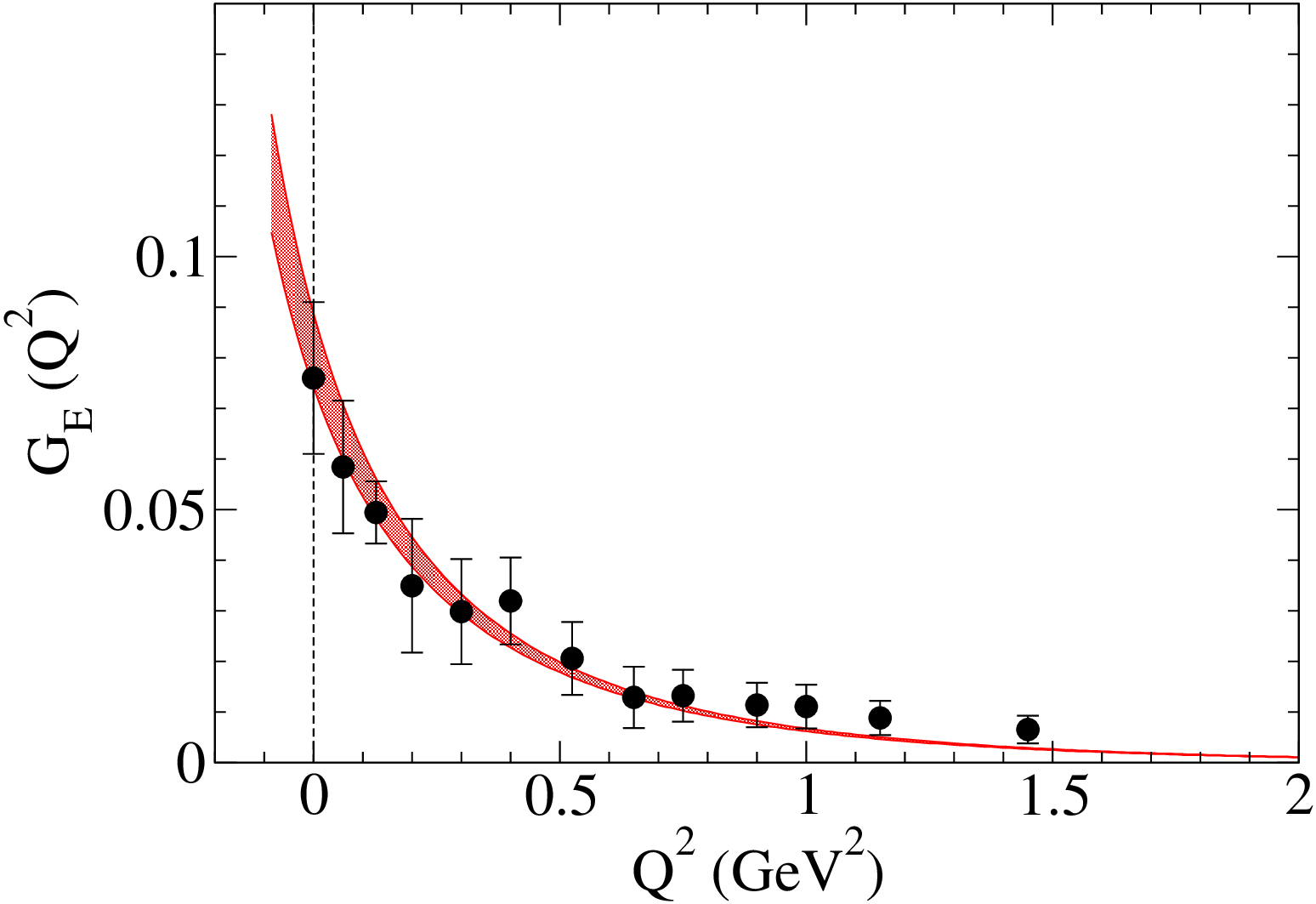}  \hspace{1.5cm}
\includegraphics[width=2.1in]{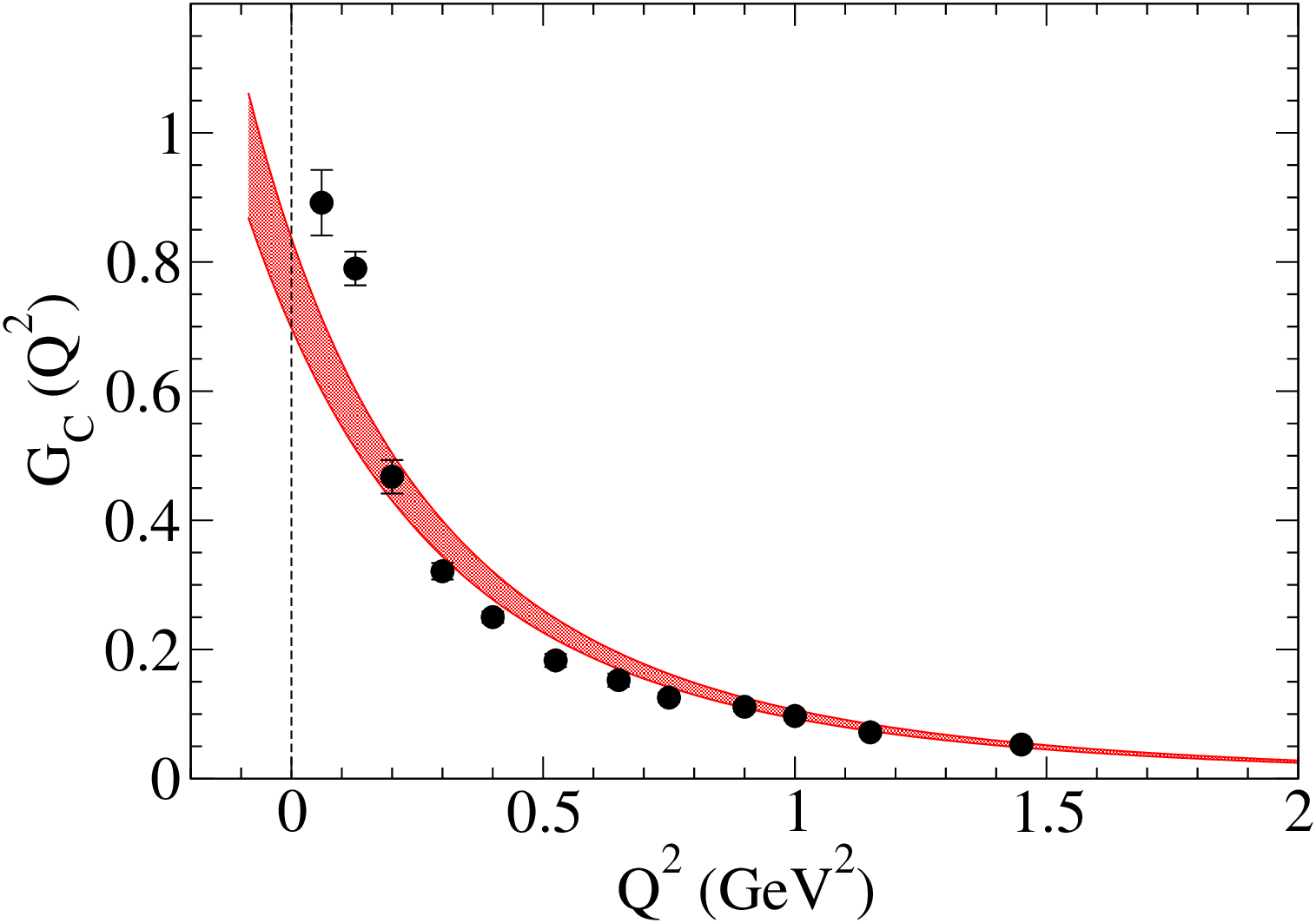}
}}
%\caption{Form factors $G_E$ and $G_C$ obtained using the improved large-$N_c$ parametrization for the pion cloud contributions (see description in the text) combined with an estimate from the valence quark contributions~\cite{SiegertQuad,LatticeD}. The bands indicate relative variation of $1/N_c^2$. Data from Ref.~\cite{MokeevDatabase}.}
\label{figD1232-2}
\end{figure*}
\vspace{-.7cm}
\noindent{\bf Fig.~4.}
{Form factors $G_E$ and $G_C$ obtained using 
the improved large-$N_c$ parametrization for 
the pion cloud contributions (see description in the text)
combined with an estimate from 
the valence quark contributions~\cite{SiegertQuad,LatticeD}.
The bands indicate relative variation of $1/N_c^2$.
Data from Ref.~\cite{MokeevDatabase}.}

\section{Summary and conclusions}

We discussed the constraints associated 
with the Siegert's theorem for the 
 $\gamma^\ast N \to N(1535)1/2^-$, $\gamma^\ast N \to N(1520)3/2^-$
and  $\gamma^\ast N \to \Delta(1232)3/2^+$ transition amplitudes.
In all cases we obtain a relation between 
the electric and scalar amplitudes given by 
$E \propto S/|{\bf q}|$ when $|{\bf q}| \to 0$.
For the  $\gamma^\ast N \to \Delta(1232)3/2^+$ transition,
the previous relation has  to be corrected by a factor of $1/|{\bf q}|$,
since the exact condition is $G_E = \kappa G_C$,
and $G_E \propto E/|{\bf q}|$, $G_C \propto S/|{\bf q}|^2$.
For the  $\gamma^\ast N \to N(1520)3/2^-$ transition
there is an extra relation between the transverse amplitudes 
at the pseudothreshold.

For all the transitions we propose parametrizations
compatible with the data and with the Siegert's theorem.
For the case of the $\gamma^\ast N \to \Delta(1232)3/2^+$ 
we discuss the constraints of the  Siegert's theorem
to the parametrizations of the pion cloud contributions 
for the quadrupole form factors.
Moreover we propose a parametrization for $G_E^\pi$
which improves the agreement with the Siegert's theorem
and gives a good description of  the empirical data.

\vspace{.5cm}
{\bf Acknowledgments:}
This work is supported by the Brazilian Ministry of Science,
Technology and Innovation %(MCTI-Brazil)
and by the project 
``Science without Borders'' from the Conselho Nacional
de Desenvolvimento Cient\'{i}fico e Tecnol\'ogico (CNPq) 400826/2014-3.

\clearpage

\end{document}